\documentclass{PHYEAUTH}
\usepackage{graphicx}
\usepackage{amsmath}
\usepackage{amssymb}
\usepackage{dcolumn}

\begin{document}

\begin{frontmatter}

\title{Electron Dynamics in Quantum Dots on Helium Surface}

\author[address1]{M.I. Dykman\thanksref{thank1}},
\author[address1]{P.M. Platzman,}
and
\author[address1]{P. Seddighrad}

\address[address1]{Department of Physics and Astronomy and the
Institute for Quantum Sciences, Michigan State University, East
Lansing, MI 48824, USA}

\thanks[thank1]{
Corresponding author.
E-mail: dykman@pa.msu.edu}

\begin{abstract}
We study single-electron quantum dots on helium surface created by
electrodes submerged into the helium. The intradot potential is
electrostatically controlled. We find the electron energy spectrum and
identify relaxation mechanisms. Strong in-plane confinement significantly
slows down electron relaxation. Energy relaxation is due primarily
to coupling to phonons in helium.  Dephasing is determined by
thermally excited ripplons and by noise from underlying
electrodes. The decay rate can be further suppressed by a magnetic
field normal to the helium surface. Slow relaxation in combination
with control over the energy spectrum make localized electrons
appealing as potential qubits of a quantum computer.

\end{abstract}

\begin{keyword}
electrons on helium \sep quantum computers  \sep quantum dots \sep relaxation mechanisms
\PACS 73.21.-b \sep 03.67.Lx
\end{keyword}
\end{frontmatter}


The system of electrons on the surface of superfluid $^4$He is
attractive from the point of view of making a scalable quantum
computer (QC) \cite{PD99}. The electrons have an extremely long
relaxation time and display the highest mobility known in a
condensed-matter system \cite{Shirahama-95}. A QC can be made by
submerging a system of individually addressed micro-electrodes beneath
the helium surface.  The typical interelectron distance is
comparatively large, $\sim 1\, \mu$m, which simplifies fabrication of
an electrode array \cite{Goodkind01}. The electrode potential, the
high barrier that prevents electrons from penetrating into the helium,
and the helium image potential together create a single-electron
quantum dot above each electrode. The parameters of the dot can be
controlled by the electrode potential.

Here we study the energy spectrum and dissipation processes for
electrons in quantum dots on helium surface. We discuss mechanisms of
coupling to helium excitations, phonons and ripplons, and the
dependence of the electron relaxation rate on the quantum dot
parameters. We investigate the role of a magnetic field normal to the
surface and of the electron-electron interaction. Decay and
decoherence of the electron states result also from classical and quantum
electrode noise. We relate the corresponding relaxation rates to the
power spectrum of the fluctuating electric field on the electron and
analyze their dependence on the parameters of the electrodes and
external leads.

\begin{figure}[h]
\begin{center}\leavevmode
\includegraphics[width=2.4in]{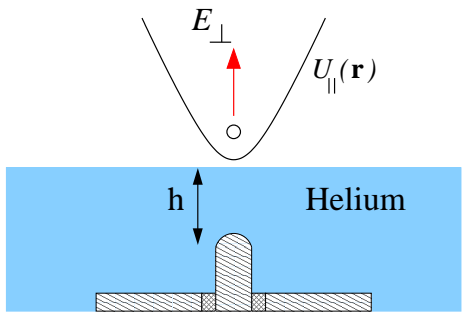}
\caption{ Geometry of a single-electron quantum dot on helium
surface. A micro-electrode is submerged by the depth $h\sim 0.5\,\mu$m
beneath the helium surface. The electron is driven by a field
$E_{\perp}$ normal to the surface. This field is a sum of the
electrode field $\mathcal{E}_{\perp}$ and the field of the
parallel-plate capacitor (only the lower plate of the capacitor is
shown). The confining in-plane potential $U_{_{\parallel}}({\bf
r})$ [${\bf r}=(x,y)$] is
determined by the electrode potential and geometry. }
\label{fig:electrode}
\end{center}
\end{figure}

The geometry of a quantum dot can be understood from
Fig.~\ref{fig:electrode}. The potential is a sum of the out-of-plane
and in-plane parts. The out-of-plane potential is similar to that in
the absence of the electrode. It leads to quantization of motion
normal to the surface. In the absence of the field $E_{\perp}$ the
energy levels are $E_n=-R/n^2$ ($n=1,2,\ldots$), where the effective
Rydberg energy $R\approx 8$~K \cite{Andrei_book}. The in-plane
potential $U_{\parallel}({\bf r})\approx
m\omega_{_{\parallel}}^2r^2/2$ [${\bf r}=(x,y)$] is parabolic, for
typical parameter values.

In the absence of a magnetic field, electron states in a dot
$|n,\nu,m_{\nu}\rangle$ are characterized by the quantum number $n$ of
motion normal to the surface, the principal quantum number $\nu$ of
vibrations about an equilibrium in-plane position, and the number
$m_{\nu}$ that enumerates degenerate vibrational states.

A confined electron can serve as a qubit \cite{PD99}. The working
states of the qubit are $|1,0,0\rangle$ and $|2,0,0\rangle$. The
energy difference between these states $E_2-E_1$ can be Stark-shifted
by the electric field from the electrode $\mathcal{ E}_{\perp}$. The
shift of 1~GHz occurs if $\mathcal{ E}_{\perp}$ is changed by $\sim
1$~V/cm. The field $\mathcal{ E}_{\perp}$ also determines the in-plane
vibrational frequency $\omega_{_{\parallel}}$. A simple estimate of
$\mathcal{ E}_{\perp}$ can be made by assuming that the electrodes are
spheres of radius $r_{\rm el}$. In this case
$\omega_{_{\parallel}}=(e\mathcal{ E}_{\perp}/mh)^{1/2}$. Typically
$\omega_{_{\parallel}}/2\pi \sim 20$~GHz, whereas the transition
frequency $\Omega_{\rm tr}=(E_2-E_1)/\hbar$ is 6-10 times larger.

Because of the discreteness of the electron energy spectrum in a dot,
the mechanisms of electron decay and dephasing are qualitatively
different from those studied for a 2D electron system on helium in the
absence of in-plane confinement \cite{Ando-85}. In particular, the
major known scattering mechanism, quasi-elastic scattering by ripplons,
does not work. Decay of the excited state $|2,0,0\rangle$ is most
likely to occur via a ripplon- or phonon-induced transition to the
closest in energy excited vibrational state of the electron
$|1,\nu_{c},m_{\nu}\rangle$, with $\nu_c={\rm
int}[(E_2-E_1)/\hbar\omega_{_{\parallel}}]$. The energy transfer in
the transition is $\delta E=E_2-E_1-\nu_c\hbar\omega_{_{\parallel}}
\sim \hbar\omega_{_{\parallel}}$. It largely exceeds the energy of
ripplons with wave numbers $q\lesssim 1/a_{_{\parallel}}$, where
$a_{_{\parallel}} =(\hbar/m\omega_{_{\parallel}})^{1/2}$ is the
in-plane electron localization length in the dot. This makes
one-ripplon decay exponentially improbable and strongly reduces the
decay rate compared to the case of electrons that are free to move
along the helium surface.

An electron in a dot can decay by emitting two ripplons that will
propagate in opposite directions with nearly same wave numbers
$q_{1,2}$, which are determined by the condition that the ripplon
frequency is $\omega_r(q_{1,2})\approx \delta E/2\hbar$ [$|{\bf q}_1+{\bf
q}_2|\lesssim 1/a_{_{\parallel}} \ll q_{1,2}$]. Alternatively, and
even with higher probability, decay may occur through an electron
transition accompanied by emission of a phonon. The appropriate
phonons propagate nearly normal to the helium surface: their wave
vectors make an angle with the normal to the surface $\sim
(mv_s^2/\hbar\omega_{_{\parallel}})^{1/2}\ll 1$, where $v_s$ is the
sound velocity in helium. We propose two mechanisms of coupling to
phonons \cite{DPS03}. One is phonon-induced deformation of the helium
surface.  The second is phonon-induced modulation of the dielectric
constant of helium and thus of the electrostatic energy of the
electron above helium.

Relatively simple expressions for the electron decay rate for
different coupling mechanisms are obtained by approximating the
electron potential at the helium surface by a sharp infinitely high
potential wall. Such an approximation applies only if the wavelength
of helium vibrations is much bigger than the width $\Delta_d$ of the
diffuse surface layer, which is of the order of a few angstroms
\cite{Penanen-Pershan00}. When we estimate the decay rate numerically,
we make an assumption that helium vibrations are essentially decoupled
from the electrons when their wave numbers exceed $\Delta_d^{-1}$.
Then, for typical in-plane electron frequencies
$\omega_{_{\parallel}}/2\pi \sim 20$~GHz, both the two-ripplon and
phonon decay rates are determined primarily by electron transitions
with the smallest energy transfer $\delta E\sim
\hbar\omega_{_{\parallel}}$. Transitions over several electron
vibrational levels, with energy transfer $n\hbar\omega_{_{\parallel}}$
with $n\gg 1$, can be disregarded. This gives the decay rate $\lesssim
10^4$~s$^{-1}$, which is presumably an overestimate.

Electron dephasing due to coupling to excitations in helium comes
primarily from quasi-elastic scattering of ripplons off the
electron. Scattering amplitudes are different in different electron
states, and therefore scattering leads to diffusion of the phase
difference between the wave functions $|2,0,0\rangle $ and
$|1,0,0\rangle$. The typical wave numbers $q_{\phi}$ of the ripplons
that contribute to dephasing are determined by the condition that the
ripplon frequency $\omega_r(q_{\phi})\lesssim k_BT/\hbar$. They are much
smaller than $1/\Delta_d$, and the coupling is well described by the
approximation of a sharp helium surface. For different mechanisms of
coupling to ripplons the dephasing rate $\Gamma^{(\phi)}$
displays same temperature dependence $\Gamma^{(\phi)} \propto
T^3$, which is much slower than the standard $T^7$ law for defects in
solids. Numerically, $\Gamma^{(\phi)} \lesssim 10^2$~s$^{-1}$ for
$T=10$~mK.

The electron decay rate may be further reduced by applying a magnetic
field normal to the helium surface. As we will show, this is
equivalent to further increasing $\omega_{_{\parallel}}$ and should
require helium vibrations with even shorter wavelengths in order to
meet the condition of energy conservation in decay. The theory will
also describe the case of many quantum dots. In this case, the energy
spectrum of in-plane electron excitations consists of plasmon bands
and is continuous. Still, as we show, the relaxation rate remains
strongly suppressed compared to the case of unconfined electrons.

The in-plane electron coordinate ${\bf r}_n$ in an $n$th dot can be
expanded in the creation and annihilation operators of the in-plane
vibrational modes of the electrons $a_{kj}, a^+_{kj}$,
\begin{equation}
\label{coordinate}
{\bf r}_n={\bf R}_n + \sum\nolimits_{kj}[{\bf A}^{(n)}_{kj}a_{kj} + {\rm H.c}].
\end{equation}
Here, ${\bf R}_n$ is the equilibrium in-plane position, and $j=1,2$
enumerates vibrational modes. The quantum number $k$ can be set equal
to zero in the case of one dot, whereas for a periodic array of dots it
becomes a plasmon wave vector ${\bf k}$.

For one dot ($n=1$) in a
strong magnetic field ${\bf B}$ antiparallel to $\hat{\bf z}$
\begin{eqnarray}
\label{frequencies}
\quad\omega_{01}\approx \omega_c=|eB|/mc,\quad \omega_{02}\approx \Omega_0
\equiv\omega_{_{\parallel}}^2/\omega_c,\\
{\bf A}
_1\approx i {\bf A}
_2\approx 2^{-1/2}l(\hat{\bf x}+i\hat{\bf y}) \qquad
[{\bf A}_j \equiv {\bf A}^{(1)}_{0j}],\nonumber
\end{eqnarray}
where $l=(\hbar/m\omega_c)^{1/2}]$ is the magnetic length, and we
assumed that $\omega_{_{\parallel}}$ is small compared to the
cyclotron frequency $\omega_c$. For a multi-dot system, the
vibrational frequencies form two bands. The bandwidths are $\lesssim
\omega_p^2/\omega_c$, where $\omega_p=(2\pi e^2/md^3)^{1/2}$ is the
typical plasma frequency ($\omega_p\ll \omega_c$) and $d$ is the
interelectron distance. They are further reduced if $\omega_p \ll
\omega_{_{\parallel}}$. The minimal frequency of the upper band is
$\approx\omega_c$, whereas that of the lower band is $ \sim\Omega_0$.

The Hamiltonian that describes $|2\rangle \to |1\rangle$ transitions
 induced by excitations in helium has the form
\begin{equation}
\label{decay_coupling}
H_i^{\rm (d)}=\sum\nolimits_n|2\rangle_n\, _n\langle 1|
\sum\nolimits_{\bf q}\hat {V}_{\bf q}e^{i{\bf qr_n}} + {\rm H.c.}.
\end{equation}
Here, $|1\rangle_n$ and $|2\rangle_n$ are the states of an $n$th
electron normal to the surface, and $\hat{V}_{\bf q}$ is the operator
that depends on the coordinates of helium vibrations, i.e., phonons
and ripplons. The wavelengths of the vibrations involved in electron
scattering are much smaller than the interelectron distance. Therefore
each electron has its ``own'' thermal bath of helium excitations. In
the Born approximation, the decay rate of the state $|2\rangle_n$ for an
$n$th electron is
\begin{eqnarray}
\label{decay_rate}
&&\Gamma_{n}^{\rm (d)}= \hbar^{-2}{\rm
Re}\,\int\nolimits_0^{\infty}dt
e^{i\Omega_{\rm tr}t}\sum\nolimits_{\bf q}
\tilde{\mathcal{S}}_n({\bf q},t) \nonumber\\
&&\;\times \langle\hat{V}_{\bf q}(t)\hat{V}_{-\bf q}(0)\rangle,
\quad \tilde{\mathcal{S}}_n({\bf q},t)
=  \langle e^{i{\bf qr}_n(t)} e^{-i{\bf qr}_n(0)} \rangle.
\end{eqnarray}
Here, the averaging $\langle\cdot\rangle$ is performed assuming that
the electron and helium vibrations are uncoupled. We have also disregarded
the difference between the transition frequencies $\Omega_{\rm
tr}=(E_2-E_1)/\hbar$ for different electrons.

From Eq.~(\ref{coordinate}), for $k_BT\ll \hbar\Omega_0$ the structure
factor $\tilde{\mathcal{S}}_n({\bf q},t)$ (\ref{decay_rate}) has the
form
\begin{eqnarray}
\label{W}
&&\tilde{\mathcal{S}}_n({\bf q},t)=\exp[-q^2W_n(t)/2],\\
&&W_n(t)=\sum\nolimits_{kj}|{\bf
A}^{(n)}_{kj}|^2[1-\exp(-i\omega_{kj}t)].\nonumber
\end{eqnarray}
The effective Debye-Waller factor $W_n(t)$ is independent of the
electron number $n$ if the coefficients ${\bf A}^{(n)}_{kj}$ for
different electrons differ only by a phase factor, as in the case of
a periodic set of dots.

In what follows we will assume that the level spacing $E_2-E_1$ is of
the same order as the distance between the Landau levels
$\hbar\omega_c$ and that the energy deficit $\delta\tilde{E}
=E_2-E_1-\tilde{\nu}_c\hbar\omega_c$ largely exceeds $\hbar\Omega_0$
[here, $\tilde{\nu}_c={\rm int}(E_2-E_1)/\hbar\omega_c$]. Then many
vibrations of the lower vibrational branch $j=2$ are excited in a
transition. This means that the structure factor
$\tilde{\mathcal{S}}_n({\bf q},t)$ can be evaluated assuming that
$\Omega_0 t\ll 1$. In doing this we will take into account the relations
$\sum\nolimits_k|{\bf A}^{(n)}_{k1}|^2 \approx l^2$ and
\begin{equation}
\label{sum_rule}
\sum_k|{\bf A}^{(n)}_{k2}|^2\omega_{k2} \approx l^2\Omega_n,
\;\Omega_n = [\partial^2U_{_{\parallel}}/\partial{\bf
r}_n^2]/2m\omega_c,
\end{equation}
where $U_{_{\parallel}}$ is the overall in-plane potential of an $n$th
electron that includes the Coulomb energy of the electron-electron
interaction, and the Laplacian is calculated at the equilibrium
position. It is seen from Eq.~(\ref{frequencies}) that
Eq.~(\ref{sum_rule}) applies in the case of one dot. A proof for a
multi-dot system will be discussed elsewhere.

From Eq.~(\ref{sum_rule}) we obtain
\begin{eqnarray}
\label{asymptotic_S}
\tilde{\mathcal{S}}_n({\bf q},t)&&\approx \exp[-(q^2l^2/2)
(1+i\Omega_nt)]\nonumber\\
&&\times\sum\nolimits_{\nu}{1\over \nu !}(q^2l^2/2)^{\nu}\,e^{-i\nu\omega_ct}.
\end{eqnarray}
This equation shows that, in an electron $|2\rangle \to |1\rangle$
transition, the energy transferred to the lower-branch vibrational
modes is $\sim q^2l^2\Omega_n\hbar$. The typical values of the factor
$q^2l^2$ are $\lesssim 1$, otherwise the transition probability
becomes exponentially small. This means that, even though the energy
spectrum of a set of dots is band-like, the typical energy that has to
be transferred to helium excitations is $\sim \delta\tilde{E}\sim
\hbar\omega_c$. For $\omega_{_{\parallel}}\ll \omega_c$ this is a much
bigger energy than in the absence of a magnetic field. Therefore we
expect that a magnetic field can significantly reduce the decay rate,
because decay requires helium excitations with wavelengths smaller
than the width of the surface diffuse layer.

The occurrence of a mode with frequency $\Omega_0\ll
\omega_{_{\parallel}}$ in a magnetic field leads to an increase of the
ripplon-induced dephasing rate. This can be seen in the higher order
in the electron-ripplon coupling where account is taken of virtual
ripplon-induced transitions between the electron vibrational levels,
cf. \cite{PD99,DPS03}. The amplitude of such transitions increases
with the decreasing $\Omega_0$, i.e., with the increasing magnetic
field. The effect imposes a limitation on the field magnitude. One
concludes that there exists an optimal range of magnetic fields where
both decay and dephasing rates are small and of the same order of
magnitude.

An important source of dephasing of electron states in a
single-electron dot is noise from the underlying electrode, see
Fig.~1. We will discuss it for one dot, as the role of inter-dot
interaction is minor in this case. Coupling to the electrode noise is
dipolar, with the Hamiltonian %
\begin{equation}
\label{dipolar_coupling}
H_{\rm dip}= e \,\delta\hat{\mathcal{E}}_{\perp} z,
\end{equation}
where $\delta\hat{\mathcal{ E}_{\perp}}$ is the normal to the surface
component of the field from quantum or classical charge density
fluctuations on the electrode.

The rates $\Gamma^{(\phi)}_{\rm el}$ and $\Gamma^{\rm (d)}_{\rm el}$ of
the electron dephasing and decay can be expressed \cite{DPS03} in
terms of the field correlation function
%
\begin{equation}
\label{correlator}
Q(\omega)= \int\nolimits_0^{\infty}dt\,e^{i\omega t}\langle \delta\hat{\mathcal{E}}_{\perp}(t)\delta\hat{\mathcal{E}}_{\perp}(0)\rangle
\end{equation}
as
\begin{eqnarray}
\label{raman_electrode}
&&\Gamma^{(\phi)}_{\rm el}= e^2(z_{22}-z_{11})^2
{\rm Re}~Q(0)/\hbar^2,\nonumber\\
&&\Gamma^{\rm (d)}_{\rm el}= 
e^2|z_{12}|^2{\rm Re}~Q(\Omega_{\rm tr})/\hbar^2.
\end{eqnarray}
Here, $z_{ij} = \langle i|z|j\rangle$, with $i,j=1,2$.

A major contribution to the dephasing rate comes from Johnson's
noise in the external leads. A simple estimate can be made by assuming
that the confining electrode is a sphere of radius $r_{\rm el}$. It
gives
\begin{eqnarray}
\label{spher_electrode_phi}
\Gamma^{(\phi)}_{\rm el}= 2k_BT_{\rm ext}\mathcal{ R}_{\rm
ext}e^2(z_{22}-z_{11})^2 r_{\rm el}^2/\hbar^2h^4,
\end{eqnarray}
where $T_{\rm ext}$ and $\mathcal{ R}_{\rm ext}$ are the lead
temperature and resistance.  For $\mathcal{ R}_{\rm ext}=25~\Omega$,
$T_{\rm ext}=1$~K, $r_{\rm el}=0.1\,\mu$m, and $h=0.5\,\mu$m we obtain
$\Gamma^{(\phi)}_{\rm el}\approx 1\times
10^4$~s$^{-1}$. Eq.~(\ref{spher_electrode_phi}) suggests how to reduce
the rate $\Gamma^{(\phi)}_{\rm el}$. The decay rate $\Gamma^{\rm (d)}_{\rm
el}$ is much less than the phonon-induced decay rate.

Eq.~(\ref{raman_electrode}) allows one also to estimate the effect of
noise from reorienting defects in the electrode. For electrodes
submerged into helium this noise should be weaker than in
semiconductor-based systems proposed for quantum computing, in
particular because it scales with the distance to the electrode as
$h^{-4}$, cf. Eq.~(\ref{spher_electrode_phi}).

In this paper we have calculated the relaxation rate for electrons in
quantum dots on the helium surface. We proposed new relaxation
mechanisms and found the dependence of the relaxation rate on the
potential of the confining electrodes and the magnetic field. It follows
from the results that the ratio of the relaxation rate to the clock
frequency of the QC based on electrons on helium, which is determined
by the electron-electron interaction, can be as small as
$10^{-4}-10^{-5}$, for typical interdot spacing $\lesssim 1\,\mu$m.

\end{document}